\documentstyle[psfig]{laa-s}
\begin{document}

\thesaurus{04.01.2, 08.01.2, 08.12.1, 08.06.3, 08.07.1}

\title {Library of high and mid-resolution spectra in the 
Ca~{\sc ii} H \& K, H$\alpha$, H$\beta$,  
Na~{\sc i} D$_{1}$, D$_{2}$, and He~{\sc i} D$_{3}$ 
line regions of F, G, K and M field stars
\thanks{Based on observations made with the Isaac Newton telescope and
the William Herschel Telescope
operated on the island of La Palma by the Royal Greenwich
 Observatory at the Spanish Observatorio del Roque de
Los Muchachos of the Instituto de Astrof\'{\i}sica de Canarias,
 and with the 2.2 m telescope of
the  Centro Astron\'{o}mico Hispano-Alem\'{a}n of
Calar Alto (Almer\'{\i}a, Spain)
operated jointly  by  the Max Planck Institut f\"{u}r Astronomie
(Heidelberg) and the Spanish Comisi\'{o}n Nacional de
Astronom\'{\i}a.} 
\thanks{The spectra of the stars listed 
in Table~\ref{tab:par} are
also available in electronic form at the CDS via anonymous 
ftp to cdsarc.u-strasbg.fr (130.79.128.5)
or via http://cdsweb.u-strasbg.fr/Abstract.html
}
}

\author{D.~Montes \inst{1} 
\and E.L. Mart\'{\i}n \inst{2}
\and M.J.~Fern\'{a}ndez-Figueroa \inst{1} 
\and M.~Cornide \inst{1} 
\and E.~De Castro \inst{1}}

\offprints{D.~Montes (dmg@ucmast.fis.ucm.es)}
                             
\institute{Departamento de Astrof\'{\i}sica,
Facultad de F\'{\i}sicas,
Universidad Complutense de Madrid, E-28040 Madrid, Spain
\and
Instituto de Astrof\'{\i}sica de Canarias, E-38200 La Laguna,
Tenerife, Spain
}

\date{Received ; accepted }

\markboth{D. Montes et al.: 
Library of high and mid-resolution spectra of F, G, K and M stars }{ }


\abstract{
In this work we present spectroscopic observations 
centered in the spectral lines most widely used as 
optical indicators of chromospheric activity  
(H$\alpha$, H$\beta$, Ca~{\sc ii} H \& K,  
and He~{\sc i} D$_{3}$) 
in a sample of F, G, K and M chromospherically inactive stars.
The spectra have been obtained with the aim of providing a library of 
high and mid-resolution spectra to be used in the application of the 
spectral subtraction technique to obtain the active-chromosphere
contribution to these lines in 
chromospherically active single and binary stars. 
This library can also be used for spectral classification purposes.
A digital version with all the spectra is available via ftp and the 
World Wide Web (WWW) in both ASCII and FITS formats
}
\keywords{Atlases
-- stars: activity  
-- stars: late-type 
-- stars: fundamental parameters
-- stars: general}


\maketitle

\section{Introduction}

Enhanced emission cores in the Ca~{\sc ii} 
H \& K, are the primary optical indicators of chromospheric activity
in late-type stars, but also the emission or the 
filling-in of the central core of other lines such as 
H$\alpha$, H$\beta$, Na~{\sc i} D$_{1}$, D$_{2}$, and He~{\sc i} D$_{3}$ 
indicate the existence of an active chromosphere in these stars.
Actually, the later mentioned lines are only in emission in a few 
very active stars, whereas in a large number of moderately active stars only 
a filling-in of the photospheric absorption is present.
To infer the chromospheric activity level a comparison with non-active 
stars is needed, for example by means of the spectral subtraction technique.
This technique provides reliable measurements of the 
the active-chromosphere contribution to these lines
(see Montes et al. 1995a, c; and references therein).
To apply this technique a large number of spectra of inactive stars 
(i.e., stars with negligible Ca~{\sc ii} H \& K emission) with different 
spectral types and luminosity classes taken 
with the same spectral resolution that of the stars under consideration
is needed.

Previously published stellar libraries
cover the optical range and extend to 
the near infrared, however they are of poor spectral resolution.
The more widely used have the following wavelength ranges and spectral 
resolutions:
Gunn \& Stryker (1983) (3130-10800 \AA,  20 and 40\AA); 
Jacoby et al. (1984) (3510-7427 \AA, 4.5 \AA);
Pickles (1985) (3600-10000 \AA, 15 \AA);
Kirkpatrick et al. (1991) (6300-9000 \AA, 8 and 18 \AA);
Silva \& Cornell (1992) (3510-8930 \AA, 11 \AA);
Torres-Dodgen \& Weaver (1993) (5800-8900 \AA, 15 \AA);
Danks \& Dennefeld (1994) (5800-10200 \AA, 4.3 \AA);
Allen \& Strong (1995) (5800-10200 \AA, 6 \AA) and 
Serote Roos et al. (1996) (4800-9000 \AA, 1.25 and 8.5 \AA).
As can be seen the higher spectral
resolution is only 1.25~\AA$\ $ (Serote Roos et al. 1996)
and 4.5 \AA$\ $ (Jacoby et al. 1984) that is 
much lower than needed in detailed spectroscopic studies of
chromospheric activity.


\begin{table*}
\caption[ ]{Summary of high-resolution observations \label{tab:obs1} }
\small
\begin{flushleft}
\begin{tabular}{c l l l c c c c c c c c c c c}
\noalign{\smallskip}
\hline
\noalign{\smallskip}
   &       &       &    & \multicolumn{2}{c}{Ca~{\sc ii} H\&K} &\ &
                          \multicolumn{2}{c}{H$\alpha$} &\ &
                          \multicolumn{2}{c}{H$\beta$} &\ &
                          \multicolumn{2}{c}{D$_{1}$, D$_{2}$, D$_{3}$}  \\
\cline{5-6}\cline{8-9}\cline{11-12}\cline{14-15}
\noalign{\smallskip}
 O & Date & Tel. & Detector & 
$\lambda$$_{i}$-$\lambda$$_{f}$ & $\delta\lambda$ & &
$\lambda$$_{i}$-$\lambda$$_{f}$ & $\delta\lambda$ & &
$\lambda$$_{i}$-$\lambda$$_{f}$ & $\delta\lambda$ & & 
$\lambda$$_{i}$-$\lambda$$_{f}$ & $\delta\lambda$    \\
\noalign{\smallskip}
\hline
\noalign{\smallskip}
 1 & Feb 1988 & 2.2m & RCA     &
3890-4009 & 0.198 & & -  & -  & & -   &   -  & & -   &  -  \\
 2 & Jul 1989 & 2.2m & RCA 006 & 
3883-4015 & 0.198 & & 6464-6719 & 0.50  & & -   &   -  & & -   &  -  \\
 3 & Dec 1992 & INT  & EEV5    & 
3840-4050 & 0.358 & & 6507-6764 & 0.45  & & 4778-4941 & 0.34 & &  -   &  - \\
 4 & Mar 1993 & 2.2m & TEK \#6 & 
3830-4018 & 0.420 & &  -    & -      & & -   &   -   & & -   &  -  \\
 5 & Jun 1995 & 2.2m & RCA \#11 & 
-         & -     & & 6510-6638 & 0.26  & & 4807-4926 & 0.26 & &  -   &  - \\ 
 6 & Sep 1995 & INT  & TEK3  & 
-         & -     & & 6452-6695 & 0.48 & & -   & -    & & 5762-6011 & 0.48 \\ 
\noalign{\smallskip}
\hline
\noalign{\smallskip}
\end{tabular}
\end{flushleft}
\end{table*}
\normalsize


\begin{table*}
\caption[ ]{Summary of mid-resolution observations \label{tab:obs2} }
\small
\begin{flushleft}
\begin{tabular}{c l l l c c c c c}
\noalign{\smallskip}
\hline
\noalign{\smallskip}
   &       &      &   &
\multicolumn{2}{c}{H$\alpha$} &\ &
\multicolumn{2}{c}{H$\alpha$ + Na~{\sc i} D$_{1}$, D$_{2}$} \\
\cline{5-6}\cline{8-9}
\noalign{\smallskip}
 O & Date & Tel. & Detector & 
$\lambda$$_{i}$-$\lambda$$_{f}$ & $\delta\lambda$ & &
$\lambda$$_{i}$-$\lambda$$_{f}$ & $\delta\lambda$ \\
\noalign{\smallskip}
\hline
\noalign{\smallskip}
 7 & Jan 1993 & WHT & TEK1  & -         & -     & & 5500-7000 & 2.90 \\
 8 & Apr 1993 & INT & EEV5  & -         & -     & & 5626-7643 & 3.16 \\
 9 & Jun 1995 & INT & TEK3  & 6430-6824 & 0.78 & & -         & -     \\
10 & Aug 1995 & INT & TEK3  & 6295-6918 & 1.06 & & -         & -     \\
11 & Nov 1995 & INT & TEK3  & 6344-6742 & 0.78 & & -         & -     \\
\noalign{\smallskip}
\hline
\noalign{\smallskip}
\end{tabular}
\end{flushleft}
\end{table*}
\normalsize

Our intent in this paper is to provide
a library of higher resolution spectra
($\leq$0.5~\AA) of F, G, and K chromospherically inactive stars 
to be used in the application of the spectral subtraction technique
in chromospherically active single and binary stars.
These spectra can also be used for spectral classification purposes 
(see Jaschek \& Jaschek 1990)
and specially for the spectral classification of chromospherically active
binary stars with composite spectra (see Strassmeier \& Fekel 1990).
In addition, we provide spectra of M-type 
stars with resolution significantly higher 
than in previous databases (Jacoby et al. 1984; Kirkpatrick et al. 1991, 1995).

We present a total of 170 spectra
centered in the spectral lines most widely used as 
optical indicators of chromospheric activity  
in a sample of 116 F, G, K and M field stars.

In Sect.~2 we report the details of our observations and data reduction.
The library is presented in Sect.~3 with comments on the behaviour of some
interesting spectral lines.

\section{Observations and data reduction}
               
The spectroscopic observations of inactive stars
presented here were carried out during several observing seasons,
from 1988 to 1995, within a program 
devoted to the study of optical activity indicators in 
chromospherically active single and binary stars
(Montes et al. 1994, 1995a, b, c, d, 1996a, b; Mart\'{\i}n \& Montes 1996).
The high  and mid-resolution spectra were obtained with three telescopes:
the 2.2 m Telescope at
the German Spanish Astronomical Observatory (CAHA) in Calar Alto
(Almer\'{\i}a, Spain), using a Coud\'{e} spectrograph with the
f/3 camera, 
the Isaac Newton Telescope (INT) and the William Herschel Telescope (WHT)
located at the Observatorio del Roque de Los Muchachos (La Palma, Spain), 
using the Intermediate Dispersion Spectrograph (IDS) with 
the cameras 500 and 235 at the INT 
and the ISIS double arm spectrograph at the WHT.

The different observational campaigns, the telescope and detector 
used and the spectral region observed in each season
are given in Tables~\ref{tab:obs1} and \ref{tab:obs2}.
We also give for each spectral region the 
wavelength range ($\lambda$$_{i}$-$\lambda$$_{f}$) covered
and the spectral resolution ($\delta\lambda$) achieved.

The spectra have been extracted using the standard
reduction procedures
in the MIDAS and IRAF packages (bias subtraction,
flat-field division, optimal extraction of the spectrum,
and wavelength calibration using arc lamps).
More details of the observations and data reduction for
the different observational seasons from 1988 to 1995 can be found
in Fern\'andez-Figueroa et al. (1994), Mart\'{\i}n et al. (1994) and 
Montes et al. (1995a, b, c, d, 1996b).

The high-resolution observations cover four spectral ranges:

\begin{enumerate}

\item The Ca~{\sc ii} H~(3968.47~\AA) \& K~(3933.67~\AA) line region, 
that also includes the H$\epsilon$~(3970.07~\AA) and in some cases 
the H$\zeta$~(3889~\AA) and H$\eta$~(3835~\AA) Balmer lines.

\item The H$\alpha$~(6562.8~\AA) line region that in some observational seasons 
also include the Li~{\sc i} 6708~\AA$\ $ line and the
Fe~{\sc i} 6663~\AA$\ $, Fe~{\sc i} 6678~\AA$\ $ and Ca~{\sc i} 6718~\AA$\ $
lines used in rotational velocity determinations
(Huisong \& Xuefu 1987).

\item The H$\beta$~(4861.32 \AA) line region.

\item The He~{\sc i} D$_{3}$~(5876 \AA) line region 
that also includes the Na~{\sc i} D$_{1}$~(5895.92~\AA) 
and D$_{2}$~(5889.95~\AA) lines.

\end{enumerate} 

We measured the resolution of our spectra using emission lines of arc lamps
taken on the same nights. Typically the  full  width   at
half maximum (FWHM) was two pixels.
The spectral resolution ($\delta\lambda$)
achieved ranges between 0.2 and 0.5 \AA$\ $ 
(R=$\lambda$/$\delta\lambda$, 25000 - 10000) depending on the observational
season (see Table~\ref{tab:obs1}).

The mid-resolution observations  ($\delta\lambda$ between 0.8 and 3~\AA)
cover, in some cases, the H$\alpha$ line region 
and in other cases the H$\alpha$ and 
Na~{\sc i} D$_{1}$ and D$_{2}$ line region (see Table~\ref{tab:obs2}).

In the H$\alpha$, H$\beta$, and Na~{\sc i} D$_{1}$, D$_{2}$, 
and He~{\sc i} D$_{3}$ line regions the spectra have been normalized
by fitting a polynome to the observed continuum.
However, in the Ca~{\sc ii} H \& K line region
it is very difficult to fit a continuum so the spectra have been normalized to
the measured flux in a 1~\AA$\ $ window centered at 3950.5~\AA.
This reference point at 3950.5~\AA$\ $ is not a real continuum, but it is a
relatively line-free region that could be used as a pseudo-continuum to
normalize all the  Ca~{\sc ii} H \& K spectra and that has been used by
Pasquini et al. (1988)
to develop a calibration procedure for converting the
observed line fluxes into absolute surface fluxes.
In the case of the mid-resolution spectra of M stars it is also 
difficult to establish a continuum, due to the presence of strong
molecular bands, so we have normalized these spectra by means of
the pseudo-continuum regions used by Mart\'{\i}n et al. (1996) located
at 6525-6550, 7030-7050, and 7540-7580. At lower wavelengths we included
other two regions near 5795 and 6150 \AA.
We plot the spectra normalized to those points in Figure~\ref{fig:mr}.
However, in the database available by ftp or WWW, we have divided the spectra 
only by the average continuum level in the region 6525-6550 \AA$\ $ in order
to preserve the observed shape.

\section{The library}

The stars included in the library
have been selected from the 
sample of lower main sequence stars studied in the Mount Wilson Observatory 
HK project (Baliunas et al. 1995 and references therein).
From this sample the slowly-rotating stars 
and the stars with the lower Ca~{\sc ii} H \& K spectrophometric index S
(normally lower than 0.2) were chosen.

Several stars not included in the sample of the HK Project have been observed,
because there are known to be inactive and slowly rotating stars and 
they were used by other 
authors in the application of the spectral subtraction technique
(see Strassmeier et al. 1990; Strassmeier \& Fekel 1990; Hall \& Ramsey 1992).  
Some visual companions of chromospherically active binaries 
have been observed simultaneously by locating both components  
of the visual par
in the slit when the spectra where taken. Some of these stars are
inactive ADS~1697~B (HD~13480) and some are  
little active ADS~16557~A (HD~218739), $\sigma$$^{1}$~CrB,
and ADS~8119~A (HD~98231) (see Table~\ref{tab:par}).

We have considered as chromospherically inactive stars, those which 
at our spectral resolution do not present any evidence of emission in the core 
of Ca~{\sc ii} H \& K lines.
We have found that some stars of the HK project
(HD~115417, HD~115383, HD~206860, HD~101501, HD~4628, HD~16160) 
with relatively low values of S index (0.2-0.3) present a small,
but measurable, emission in our Ca~{\sc ii} H \& K spectra
(see Montes et al. 1995c). 
Hence, they have not been used as reference stars.

Table~\ref{tab:par} presents information about the observed stars.
In this table we give
the HD and HR numbers, name, spectral type and luminosity class (T$_{\rm sp}$), 
from the Bright Star Catalogue 
(Hoffleit \& Jaschek 1982; Hoffleit \& Warren 1991), 
the Catalogue of Nearby Stars (Gliese \& Jahreiss 1991), and
Kirkpatrick et al. (1995),
metallicity [Fe/H] (from Taylor 1994; 1995),
rotational period (P$_{\rm rot}$) and {\it v}~sin{\it i} 
(from Donahue 1993; Baliunas et al. 1995). 
The T$_{\rm sp}$ given between brackets are from  Hoffleit \& Warren (1991) 
and the values of {\it v}~sin{\it i} marked with "*" are from the 
references given in Strassmeier \& Fekel (1990).
We also give the Ca~{\sc ii} H \& K spectrophometric index S
from Baliunas et al. (1995) or from Duncan et al. (1991) (values with "*").
In the columns labeled with   
H$\alpha$,  He~{\sc i}~D$_{3}$,  H$\beta$, and  Ca~{\sc ii}
we list information about the observations
for each spectral range, using a code given 
in the first column of  Tables~\ref{tab:obs1} and ~\ref{tab:obs2}.
In the last column ``A" and ``R" mean active and reference star respectively, 
according with our above mentioned criterion, 
and ``E" means that the H$\alpha$ line is in emission in our spectra.
In some cases, we have available spectra of several stars that have been 
classified with the same spectral type, these spectra present small 
differences in the lines that could be attibuted 
to differences in metallicities, rotation,
errors in the spectral classification, or even to variations
in the small level of activity that these stars could present.

\begin{figure*}
\psscalefirst
\vspace{-0.85cm}
\hspace{-1.6cm}
{\psfig{figure=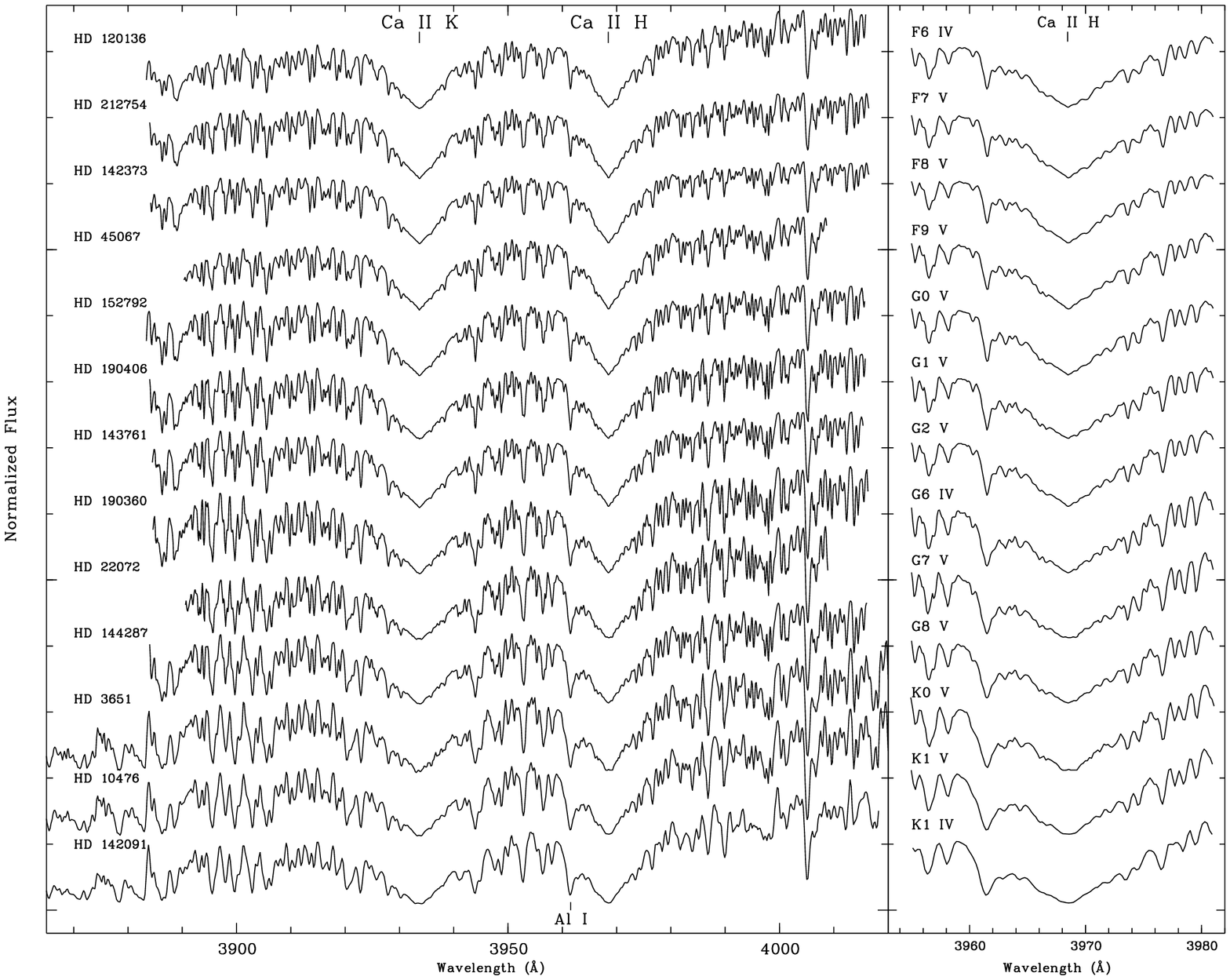,height=37.5cm,width=15.1cm,rheight=25.5cm,rwidth=6cm,angle=270}}
\caption[ ]{High-resolution spectra in the
Ca~{\sc ii} H \& K line region
\label{fig:caii}}
\end{figure*}

\begin{figure*}
\psscalefirst
\vspace{-0.85cm}
\hspace{-1.6cm}
{\psfig{figure=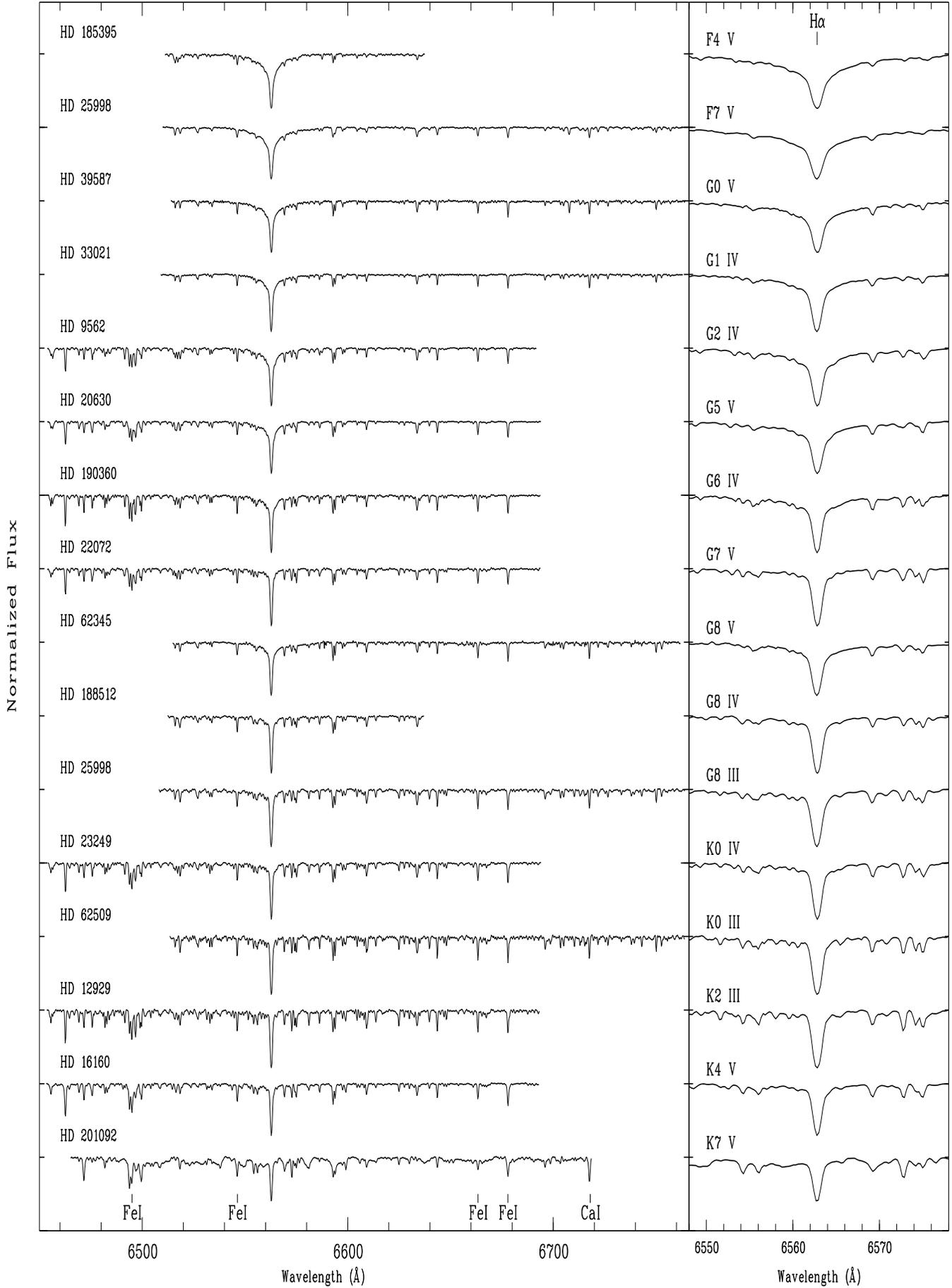,height=37.5cm,width=15.1cm,rheight=25.5cm,rwidth=6cm,angle=270}}
\caption[ ]{High-resolution spectra in the H$\alpha$ line region
\label{fig:ha}}
\end{figure*}

\begin{figure*}
\psscalefirst
\vspace{-0.85cm}
\hspace{-1.6cm}
{\psfig{figure=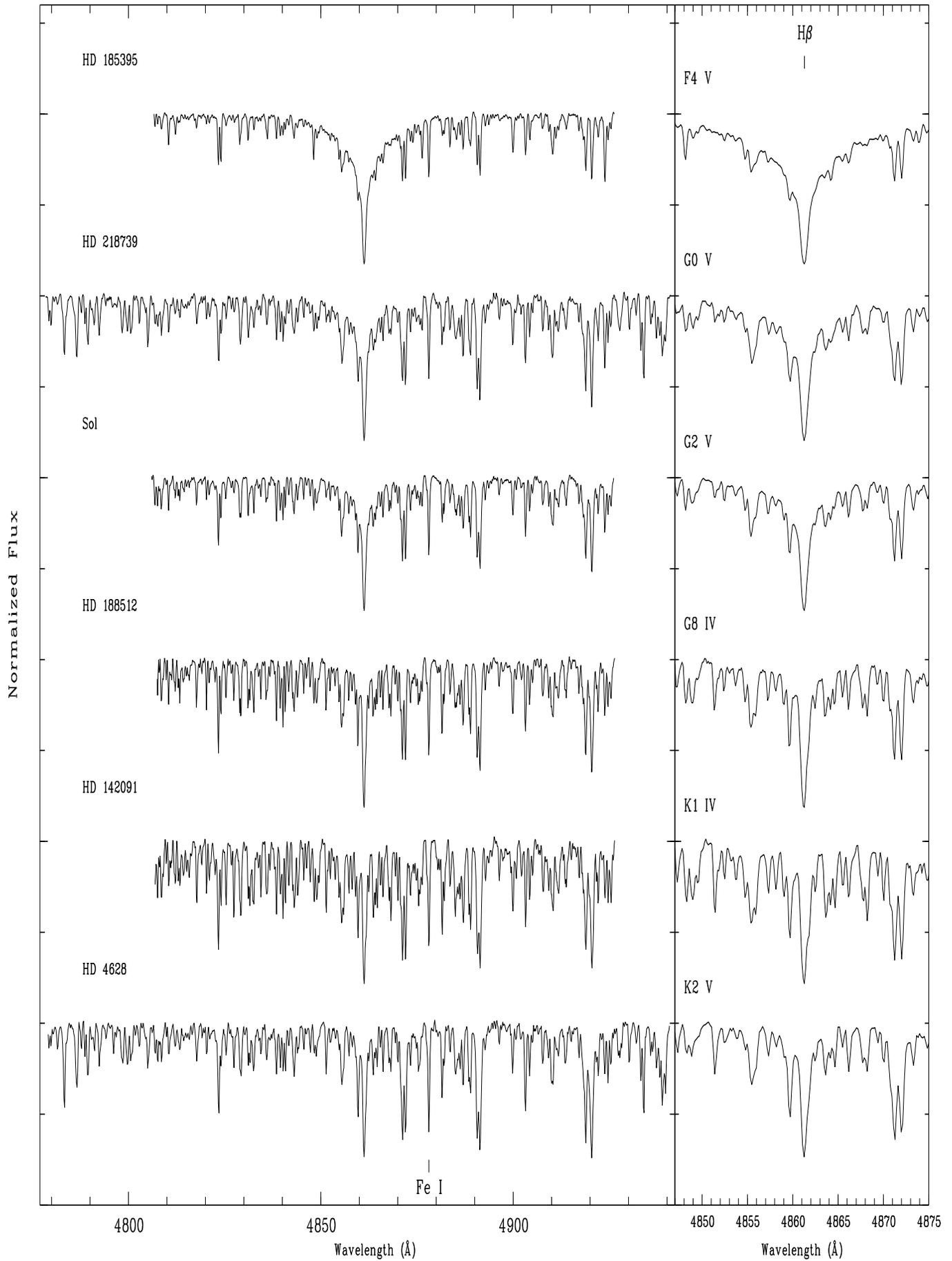,height=37.5cm,width=15.1cm,rheight=25.5cm,rwidth=6cm,angle=270}}
\caption[ ]{High-resolution spectra in the H$\beta$ line region
\label{fig:hb}}
\end{figure*}

\begin{figure*}
\psscalefirst
\vspace{-0.85cm}
\hspace{-1.6cm}
{\psfig{figure=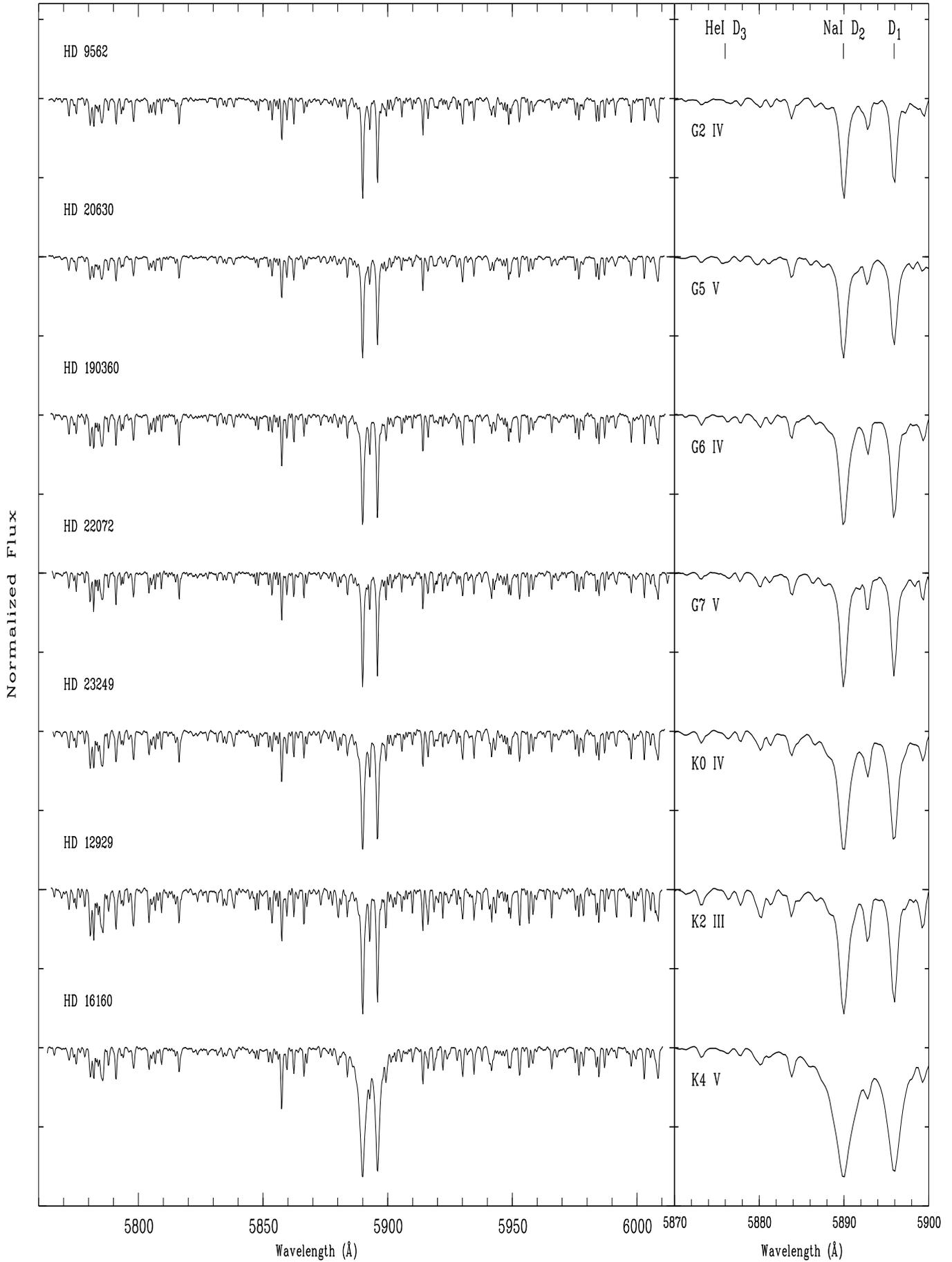,height=37.5cm,width=15.1cm,rheight=25.5cm,rwidth=6cm,angle=270}}
\caption[ ]{High-resolution spectra in the
Na~{\sc i} D$_{1}$, D$_{2}$, and He~{\sc i} D$_{3}$ line region
\label{fig:hei}}
\end{figure*}

\begin{figure*}
\psscalefirst
\vspace{-0.85cm}
\hspace{-1.6cm}
{\psfig{figure=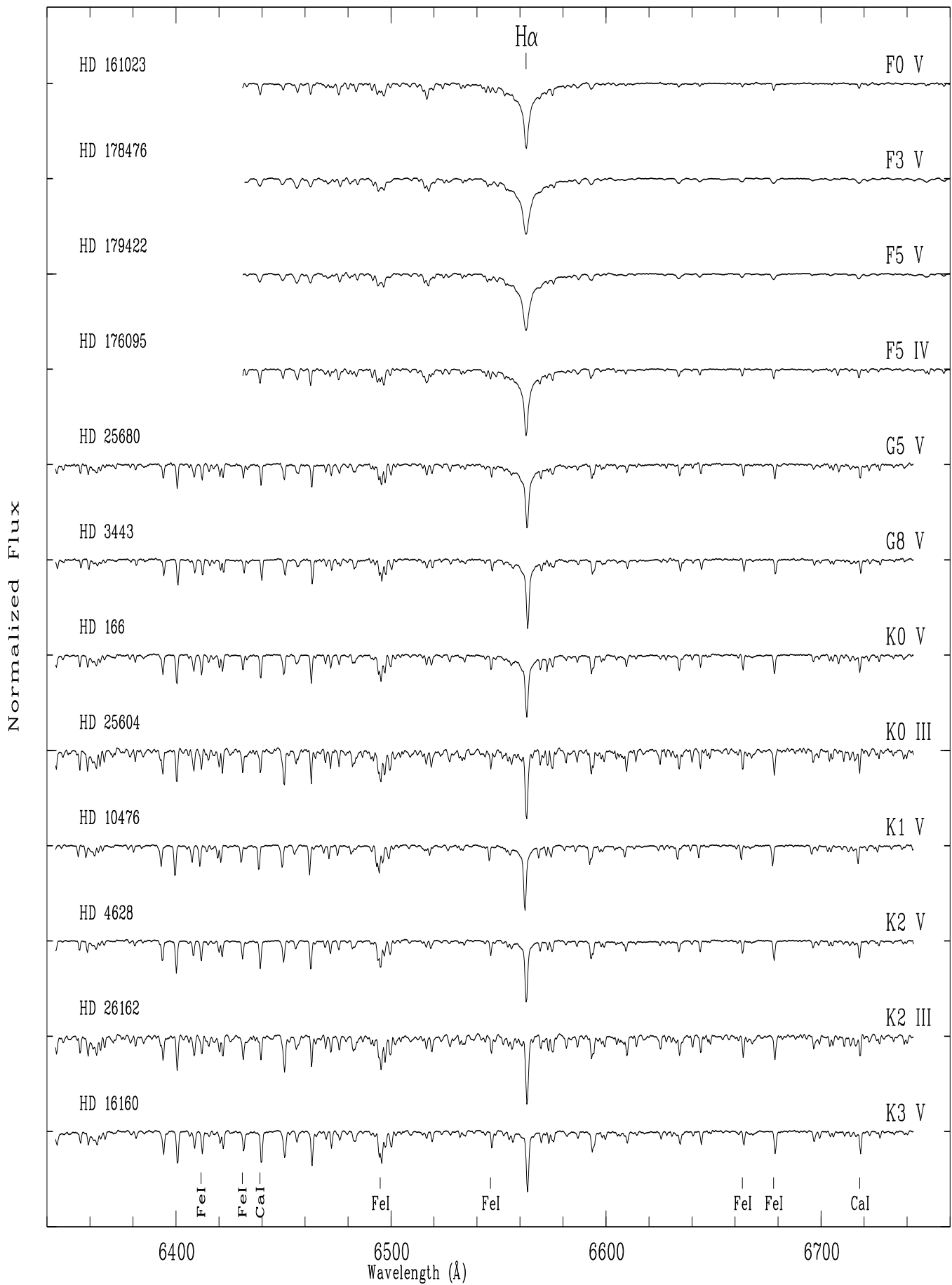,height=37.5cm,width=15.1cm,rheight=25.5cm,rwidth=6cm,angle=270}}
\caption[ ]{Mid-resolution spectra in the H$\alpha$ line region 
in the wavelength range 6340 to 6740 \AA.
\label{fig:hamr}}
\end{figure*}

Figures~\ref{fig:caii}, \ref{fig:ha}, \ref{fig:hb}, and \ref{fig:hei} show 
representative high-resolution spectra in each spectral range. 
In these figures we plot, at the left, the complete wavelength range covered 
in each spectrum.
For a better display of the spectral features
an small region of 30 \AA$\ $ centered in the spectral line 
of interest in each case is showed at the right.
Figure~\ref{fig:hamr} presents mid-resolution spectra centered 
in the H$\alpha$ line from 6340 to 6740 \AA, and  
Figure~\ref{fig:mr} shows representative mid-resolution spectra
in the wavelength range 5700 to 7600 \AA$\ $ which include the 
Na~{\sc i} D$_{1}$, D$_{2}$ and  H$\alpha$ lines. 
The stars in these figures are arranged in order of spectral type from F to M.
The HD number and the spectral type of the stars are given in each spectrum.


\begin{table*}
\caption[ ]{Stellar parameters and spectral region observed
\label{tab:par} }
\begin{flushleft}
\scriptsize
\begin{tabular}{l l l l r c c l l l l l l}
\hline
\noalign{\smallskip}
 HD & HR & Name &  T$_{\rm sp}$ & [Fe/H] & P$_{\rm rot}$ & {\it v}~sin{\it i} & S 
& H$\alpha$ & He~I~D$_{3}$  & H$\beta$ & Ca~II & A/R \\
    &    &      &       & (dex)   & (days)   & (km s$^{-1}$) &
& &  &   & &       \\
\hline
 {\bf F stars}  \\
\hline
\noalign{\smallskip}
 161023 & 6600 & -            & F0V  & -    & -    & $<$ 15& -     & 9   &   &   &   &   \\
 177552 & 7231 & -            & F1V  & -    & -    & 45    & -     & 9   &   &   &   &   \\
 178476 & 7363 & -            & F3V  & 0.170& -    & 50    & -     & 9   &   &   &   &   \\
 185395 & 7469 & $\theta$ Cyg & F4V  & 0.009& -    & 3.4*  & -     & 5   &   & 5 &   &   \\
 13480B & 642B & 6 Tri B   &    F5V  & -    & 2.236& -     & -     &     &   &   & 3 & R \\
 179422 & 7280 & -            & F5V  & -    & -    & 40    & -     & 9   &   &   &   &   \\
 176095 & 7163 & -            & F5IV & -    & -    & $<$ 10& 0.202 & 9   &   &   &   &   \\
 120136 & 5185 & $\tau$ Boo&   F6IV  & 0.096& -    & 10.0  & 0.191 &     &   &   & 2 & R \\
  82328 & 3775 & $\theta$ UMa& F6IV  &-0.172& -    & 6.4*  & 0.182*& 3   &   &   &   &   \\  
 124850 & 5338 & $\iota$ Vir&  F6III &-0.129& -    & 15.0  & 0.210 &     &   &   & 2 & R \\
 187013 & 7534 & 17 Cyg  &     F7V   &-0.109& -    & 10.0  & 0.154 &     &   &   & 2 & R \\
 212754 & 8548 & 34 Peg  &     F7V   &-0.061& -    & 10.0  & 0.140 &     &   &   & 2 & R \\
  25998 & 1278 & 50 Per  &     F7V   & -    & 2.6  & 20.0  & 0.300 & 3   &   &   &   &   \\   
 216385 & 8697 & $\sigma$ Peg& F7IV  &-0.297& -    & 10.0  & 0.142 &     &   &   & 2 & R \\
 167588 & 6831 & -           & F8V   & -    & -    & $<$ 6 & -     & 10   &   &   &   & - \\
 6920   & 340  & 44 And &      F8V   &-0.230& 15.3 & $<$ 15& 0.194 &     &   &   & 3 & R \\
 45067  & 2313 &  -       &    F8V   & -    & -    & $<$ 15& 0.141*&     &   &   & 1 & R \\
 107213 & 4688 & 9 Com  &      F8V   & 0.154& -    & 10.0  & 0.135 &     &   &   & 1 & R \\
 142373 & 5914 & $\chi$ Her&   F8V   &-0.431& -    & 10.0  & 0.147 & 2   &   &   & 2 & R \\
 187691 & 7560 & o Aql   &     F8V   & 0.059& -    & 5.0   & 0.148 &     &   &   & 2 & R \\
 194012 & 7793 &  -       &    F8V   & -    & -    & 5.0   & 0.198 & 2   &   &   & 2 & R \\
 136202 & 5694 & 5 Ser  &   F8III-IV &-0.075& -    & 5.0   & 0.140 & 2   &   &   & 2 & R \\
 154417 & 6349 & V2213 Oph& F8.5IV-V & 0.099& 7.78 & 5.0   & 0.269 & 2   &   &   & 2 & A \\
 43587  & 2251 & -   &         F9V   & -    & -    & 5.0   & 0.156*&     &   &   & 1 & R \\
 78366  & 3625 & -   &         F9V   & -    & 9.67 & 5.0   & 0.248 & 8   &   &   &   & - \\
\noalign{\smallskip}
\hline
{\bf G stars} \\
\hline
\noalign{\smallskip}
115383  & 5011 & 59 Vir      & G0V   & 0.130& 3.33  & 5.0    & 0.313 &     &   &   & 2 & A \\
152792  & -    & -           & G0V   &-0.462& -     & -      & -     &     &   &   & 2 & R \\
114710  & 4983 & $\beta$ Com& G0V (F9.5V)&0.135& 12.35 &3.9*& 0.201 &     &   &   & 2 & R \\
206860  & 8314 & HN Peg      & G0V   & -    & 4.86  & 10.0   & 0.330 & 2   &   &   & 2 & A \\
29645   & 1489 & -           & G0V   & 0.074& -     & $<$  15& 0.140 &     &   &   & 1 & R \\
13974   & 660  & $\delta$ Tri& G0V (G0.5V)&-0.444& -& 10.0   & 0.232*&     &   &   & 3 & R \\
98231   & 4375 & $\xi$ UMa A & G0V   &-0.352& -     & $<$  15& -     &     &   &   & 1 & A \\
218739  & -    & ADS 16557 A & G0V   & -    & -     & -      & 0.294*&     &   & 3 & 3 & A \\
39587   & 2047 & $\chi^{1}$ Ori& G0V &-0.084& 5.36  & 10.0   & 0.325 & 3   &   &   &   &   \\ 
13421   & 635  & 64 Cet      & G0IV  & -    & -     & $<$ 15 & 0.131 &     &   &   &1,3& R \\
190406  & 7672 & 15 Sge      & G1V   & -    & 13.94 & 5.0    & 0.194 & 2   &   &   & 2 & R \\
146362  & 6064 & $\sigma$$^{1}$ CrB &G1V&-& -  & $<$ 30 & 0.264*&    &   &   & 1 & A \\ 
33021   & 1662 & 13 Ori      & G1IV  & -    & -     & 5.0    & -     & 3   &   &   &   &   \\ 
-       & -    & Sun         & G2V   &      & 25.72 & $<$ 1.7& 0.179 &     &   & 5 & 4 & R \\
143761  & 5968 & $\rho$ CrB  & G2V (G0V)&-0.185& -  & 5.0    & 0.150 & 2   &   & 5 & 2 & R \\
81809   & 3750 &  -          & G2V   &-0.319& 40.20 & 10.0   & 0.172 &     &   &   & 1 & R \\
9562    & 448  &  -          & G2IV  & 0.147& -     &  $<$ 15& 0.136 & 6   & 6 &   & 3 & R \\
12235   & 582  & 112 Psc     & G2IV  & -    & -     &  $<$ 15& 0.160 &     &   &   & 3 & R \\
217014  & 8729 & 51 Peg      & G2.5IV&-9.000& -     &  2     & 0.149 &     &   &   & 2 & R \\
186427  & 7504 & 16 Cyg B    & G2.5V &-0.002 & -     &  3.0   & 0.145*& 8   &   &   &   & - \\
159222  & 6538 & -           & G5V   & -    & -     &  -     & 0.164*& 10   &   &   &   & - \\
20630   & 996  & $\kappa^{1}$ Cet& G5V&0.133& 9.24 & 5.6*    & 0.366 & 6   & 6 &   & 3 & A \\
25680   & 1262 & 39 Tau      & G5V   & -    & -     &  3.0   & 0.281*& 11  &   &   &   & - \\
68255   & 3210 & 16 Cnc C    & G5V   & -    & -     &  -     & -     & 8   &   &   &   & - \\
78715   & 3640 & 79 Cnc      & G5III & -    & -     &  -     & -     & 8   &   &   &   & - \\
115617  & 5019 & 61 Vir      & G6V   & 0.032& -     & 2.0*   & 0.162 &     &   &   & 1 & R \\
190360  & 7670 & -           & G6IV+M6V&-9.000& -   & -      & 0.146 & 2,6 & 6 &   & 2 & R \\
3443    & 159  & -           & G8V   &-0.101& -     & 2:     & 0.183 & 11  &   &   &   & - \\
182488  & 7368 & -           & G8V   & -    & -     & -      & 0.155*&     &   &   & 2 & R \\
131156 A & 5544 A& $\xi$ Boo A& G8V  &-0.151& 6.31  & 3      & 0.461 & 2   &   &   & 2 & A \\
144287  & -    & GJ609.2     & G8V   & -    & -     & -      & 0.156*& 2   &   &   & 2 & R \\
101501  & 4496 & 61 UMa      & G8V   &-0.070& 16.68 & $<$  15& 0.311 & 3,8 &   &   & 4 & A \\
182572  & 7373 & 31 Aql      & G8IV  & -    & -     & $<$  15& 0.148 &     &   &   & 2 & R \\
188512  & 7602 & $\beta$ Aql & G8IV  & -    & -     & 2.6     & 0.136 & 5   &   & 5 & 2 & R \\
158614  & 6516 & -           & G8IV (G9IV-V)&0.056& -  & - & 0.158 &     &   &   & 2 & R \\
 62345  & 2985 & $\kappa$ Gem& G8III & -    & -     & 6*     & 0.123*& 3   &   &   &   &   \\
107383  & 4697 & 11 Com      & G8III & -    & -     & $<$ 19 & -     & 8   &   &   &   & - \\
215665  & 8667 &$\lambda$ Peg& G8IIIa& -    & -     & $<$ 19 & 0.108 & 2   &   &   & 2 & R \\
104979  & 4608 & o Vir       & G8IIIa& -    & -     & $<$ 19 & -     &     &   &   & 4 & R \\
58368   & -    & -           & G8IIIb& -    & -     & -      & -     &     &   &   & 4 & R \\
218356  & 8796 & 56 Peg      & G8Ib  & -    & -     & $<$ 17 & 0.686 & 2   &   &   & 2 & A \\
199939  & -    & -           & G9III & -    & -     & -      & -     & 2   &   &   & 2 & A \\
201657  & -    & -           & G9III & -    & -     & -      & -     & 2   &   &   & 2 & A \\
101013  & 4474 & -           & G9III & -    & -     & -      & -     &     &   &   & 4 & R \\  
\noalign{\smallskip}
\hline
\end{tabular}
\end{flushleft}
\end{table*}
\begin{table*}
\addtocounter{table}{-1}
\caption[ ]{Continue}
\begin{flushleft}
\scriptsize
\begin{tabular}{l l l l r c c l l l l l l}
\hline
\noalign{\smallskip}
 HD & HR & Name &  T$_{\rm sp}$ & [Fe/H] & P$_{\rm rot}$ & {\it v}~sin{\it i} & S 
& H$\alpha$ & He~I~D$_{3}$  & H$\beta$ & Ca~II & A/R \\
    &    &      &       & (dex)   & (days)   & (km s$^{-1}$) &
& &  &   & &       \\
\noalign{\smallskip}
\hline
 {\bf K stars} \\
\hline
\noalign{\smallskip}
166     & 8    & -           & K0V    & -    & -      & -     & 0.486*& 11  &   &   &   & - \\
3651    & 166  & 54 Psc      & K0V    &-9.000& 48.00  & -     & 0.176 &     &   &   & 3 & R \\
185144  & 7462 & 61 Dra      & K0V    &-0.045& -      & 1.5*  & 0.215 & 8   &   &   &   & - \\
23249   & 1136 & $\delta$ Eri& K0IV   & -    & -      & 2.2*  & 0.137 & 6   & 6 &   &   &   \\
45410   & 2331 & 6 Lyn       & K0III-IV&-    & -      & -     & 0.127*& 3   &   &   &   &   \\
25604   & 1256 & 37 Tau      & K0III  & -    & -      & -     & 0.105*& 10,11&   &   &   & - \\
62509   & 2990 & $\beta$ Gem & K0III  & -    & -      & 2.5*  & 0.140*& 3   &   &   &   &   \\
109345  & 4784 & -           & K0III  & -    & -      &  -    & -     & 8   &   &   &   & - \\
139195  & 5802 & 16 Ser      & K0III  & -    & -      &$<$ 17 & -     &     &   &   & 4 & R \\ 
164349  & 6713 & 93 Her      & K0.5IIb & -   & -      & $<$ 17& -     & 2   &   &   & 2 & A \\
190404  & -    & GJ 778      & K1V    &-0.087& -      & -     & 0.174*& 2   &   &   & 2 & A \\
10476   & 493  & 107 Psc     & K1V    &-0.123& 35.2 & $<$  20 & 0.198 & 11  &   &   & 3 & R \\
22072   & 1085 & -           & K1IV (G7V)& - & -      & -     & 0.131 & 6   & 6 &   & 1 & R \\
142091  & 5901 & $\kappa$ CrB& K1IV  & -     & -      & 4.5*  & -     & 5   &   & 5 & 4 & R \\
95345   & 4291 & 58 Leo      & K1III & -     & -      & $<$ 19& -     &     &   &   & 4 & R \\
163770  & 6695 & $\theta$ Her& K1IIa & -     & -      & $<$ 19& -     & 2   &   &   & 2 & A \\
22049   & 1084 & $\epsilon$ Eri& K2V &-0.165 & 11.68  & $<$ 15& 0.496 &     &   &   & 3 & A \\
4628    & 222  & -           & K2V   &-0.235 & 38.5   & -     & 0.230 & 11  &   & 3 & 3 & A \\
166620  & 6806 & -           & K2V   &-0.114 & 42.4   & 2.5*  & 0.190 & 8   &   &   &   & - \\
12929   & 617  & $\alpha$ Ari& K2III & -     & -      & $<$ 17& 0.118*& 6   & 6 &   &   &   \\ 
26162   & 1283 & 43 Tau      & K2III & -     & -      &  -    & -     & 11  &   &   &   & - \\
190608  & 7679 & 16 Sge      & K2III & -     & -      & $<$ 19& -     & 8   &   &   &   & - \\
206778  & 8308 &$\epsilon$ Peg&K2Ib  & -     & -      & $<$ 17& 0.330 & 2   &   &   & 2 & A \\
16160   & 753  & -           & K3V   &-0.297 & 48.0   & -     & 0.226 & 6,11& 6 &   & 3 & A \\
219134  & 8832 & -           & K3V   &-9.000 & -      & -     & 0.230*&     &   &   & 2 & A \\
115404  & -    & GJ 505A     & K3V (K1V)& -  & 18.47  & -     & 0.535 &     &   &   & 2 & A \\
127665  & 5429 & $\rho$ Boo  & K3III &0.183  & -      & $<$ 15& -     &     &   &   & 4 & A \\
131156 B & 5544 B&  $\xi$ Boo B&K4V  & -     & 12.28  & 20    & 1.381 &     &   &   & 2 & A \\
131873  & 5563 & $\beta$ UMi & K4III & -     & -      & $<$ 17& -     & 2   &   &   & 2 & A \\
201091 & 8085 & 61 Cyg A     &  K5V  & -     & 35.37  & 10    & 0.658 & 2,8 &   &   & 2 & A \\
201092 & 8086 & 61 Cyg B     &  K7V  & -     & 37.84  & $<$ 25& 0.986 & 2,8 &   &   & 2 & A \\
\noalign{\smallskip}
\hline
 {\bf M stars} \\
\hline
\noalign{\smallskip}
79210  & -    & GJ 338 A, LHS 260& M0V    & -    & -     &  -     & 2.113 & 8   &   &   &   & - \\
79211  & -    & GJ 338 B, LHS 261& M1III  & -    & -     &  -     & 1.955 & 8   &   &   &   & - \\
331161 & -    & GJ 767 A, LHS 3482&M0.5V  & -    & -     &  -     & -     & 8   &   &   &   & - \\
189319 & 7635 & 12 Sge            &M0III  & -    & -     &  $<$ 17& 0.254 & 8   &   &   &   & - \\
 -     & -    & GJ 767 B, LHS 3483&M2V    & -    & -     &  -     & -     & 8   &   &   &   & - \\
190658 & 7680 & -                & M2.5III& -    & -     &  -     & -     & 8   &   &   &   & - \\
 -     & -    & GJ 569 A         & M3V    & -    & -     &  -     & -     & 8   &   &   &   & E \\
189577 & 7645 & 13 Sge           & M4IIIa & -    & -     &  -     & -     & 8   &   &   &   & - \\
 -     & -    & GJ 402, LHS 294  & M4V M5V& -    & -     &  -     & -     & 8   &   &   &   & - \\
 -     & -    & GJ 406, LHS 36   & M6V    & -    & -     &  -     & -     & 8   &   &   &   & E \\
 -     & -    & GJ 1111, LHS 248 & M6.5V  & -    & -     &  -     & -     & 7   &   &   &   & E \\
 -     & -    & LHS 2243         & M8V    & -    & -     &  -     & -     & 7   &   &   &   & E \\
84748  & 3882 & R Leo            & M8IIIe & -    & -     &  -     & -     & 8   &   &   &   & - \\
 -     & -    & GJ 569 B         & M8.5V  & -    & -     &  -     & -     & 8   &   &   &   & - \\
 -     & -    & LHS 2065         & M9V    & -    & -     &  -     & -     & 7   &   &   &   & E \\
 -     & -    & LHS 2924         & M9V    & -    & -     &  -     & -     & 7   &   &   &   & E \\
\noalign{\smallskip}
\hline
\end{tabular}
\end{flushleft}
\end{table*}


Looking at 
Figures~\ref{fig:caii}, \ref{fig:ha}, \ref{fig:hb}, and \ref{fig:hei}
some conclusions concerning the behaviour of different 
spectral lines present in each spectral region can be obtained.

In the Ca~{\sc ii} H \& K line region, 
 we can note the effect of the spectral type:
the equivalent width of several metallic lines 
increases with decreasing temperature,
in particular the Al~{\sc i} 3961.52~\AA$\ $ line (see Figure~\ref{fig:caii}).

In the case of the H$\alpha$ line, we note the increasing line wings with
hotter spectral type.
At spectral type F the line exhibits extended wings that 
decrease  with decreasing temperature. The line becomes sharper at 
spectral type K (see Figure~\ref{fig:ha}).
Some strong absorption lines in this spectral region,
that could be used for radial and rotational velocity determinations are: 
the Fe~{\sc i} 6495~\AA, 6546.25~\AA, 6663.4~\AA, and 6677.9~\AA$\ $ lines
and the Ca~{\sc i} 6718~\AA$\ $ line.
The intensity of these lines increases toward later spectral types, 
in particular the H$\alpha$, Fe~{\sc i} 6495~\AA$\ $ ratio 
has been used as a spectral classification criterion 
(Danks \& Dennefeld 1994).

The H$\beta$ line presents a marked 
temperature effect in the wings (see Figure~\ref{fig:hb}) in the same way
as the H$\alpha$ line.
The FeI 4878.08~\AA$\ $ line is an isolated and strong absorption line in this 
spectral region that could be used for radial and rotational velocity 
determinations.

The He~{\sc i} D$_{3}$ line region also includes the 
Na~{\sc i} D$_{1}$  and D$_{2}$ lines, which are well known temperature and 
luminosity discriminants among late-type stars, and
they show the expected trend of metallic-line intensity increasing 
with decreasing temperature 
(O'Connell 1973; Torres-Dodgen \& Weaver 1993; 
Danks \& Dennefeld 1994; Serote Roos et al. 1996).
The effect is more important in the later spectral types and especially in the 
wing of the lines (see Figure~\ref{fig:hei}).
The behaviour of these lines confirms the spectral classification of the star
HD~22072 as G7V (Baliunas et al. 1995) rather than the K1V given by 
Hoffleit \& Jaschek (1982).
These Na~{\sc i} resonance lines are collisionally-controlled in the
atmospheres of late-type stars and
have been observed in emission or filled-in
in very active red dwarf flare stars 
(Pettersen et al. 1984: Pettersen 1989),
so the spectra of the inactive stars presented here can be used to apply 
the spectral subtraction technique to other active stars and obtain
information about chromospheric emission 
in these lines (see Montes et al. 1996b).

In the mid-resolution spectra (Figures~\ref{fig:hamr} and \ref{fig:mr}) 
in addition to the Na~{\sc i} D$_{1}$, D$_{2}$, H$\alpha$ 
and the other lines above described, 
we can also see other interesting features such as 
Fe~{\sc i} 6411.66~\AA, Fe~{\sc i} 6430.85~\AA, and Ca~{\sc i} 6439.08~\AA$\ $ 
normally used for the application of the Doppler imaging technique
(see Figure~\ref{fig:hamr}) and 
the Ca~{\sc i} (6122 and 6162~\AA) lines which are very weak 
at spectral type F and increase in strength with decreasing temperature 
(see Figure~\ref{fig:mr}).
From mid K through M stars we can also see absorption molecular bands 
of TiO in the following regions 
(5847-6058), (6090-6390), (6651-6852), (7053-7270)
and of CaH in (6346, 6482, 6389) and (6750-7050) 
(see the K and M stars in Figure~\ref{fig:mr}).
These molecular bands become very strong at the later M spectral types,
and dominate the spectrum of these stars.
For spectral type M7 or later the VO absorption band (7400-7510) is also 
present. This feature can be used as an additional spectral classifier
in the later spectral types, because it is strongly 
dependent on temperature (Kirkpatrick et al. 1995).
Finally, we note in Figure~\ref{fig:mr} the strong telluric line O$_{2}$ 
(6867 \AA), and the very deep atmospheric B-band absorption 
feature at 7600 \AA.

In order to enable other investigators to make use of the spectra of this
library, all the spectra of the stars listed 
in Table~\ref{tab:par} are available
as FITS and ASCII format files at the CDS in Strasbourg, France,
via anonymous ftp to cdsarc.u-strasbg.fr (130.79.128.5).
They are also available via the World Wide Web at:
\newline
http://www.ucm.es/OTROS/Astrof/fgkmsl/fgkmsl.html


{\it Acknowledgements}

This research has made use of the SIMBAD data base, operated at CDS, 
Strasbourg, France.
We wish to thank the staff of Calar Alto and La Palma observatories
for their efficient assistance. We have made use of the La Palma
ING data archive for retrieving some spectra.
This work has been supported by the Universidad Complutense de Madrid
and the Spanish Direcci\'{o}n General de Investigaci\'{o}n
Cient\'{\i}fica y  T\'{e}cnica (DGICYT) under grant PB94-0263.





\newpage


\begin{figure*}
\psscalefirst
\vspace{-0.85cm}
\hspace{-1.6cm}
{\psfig{figure=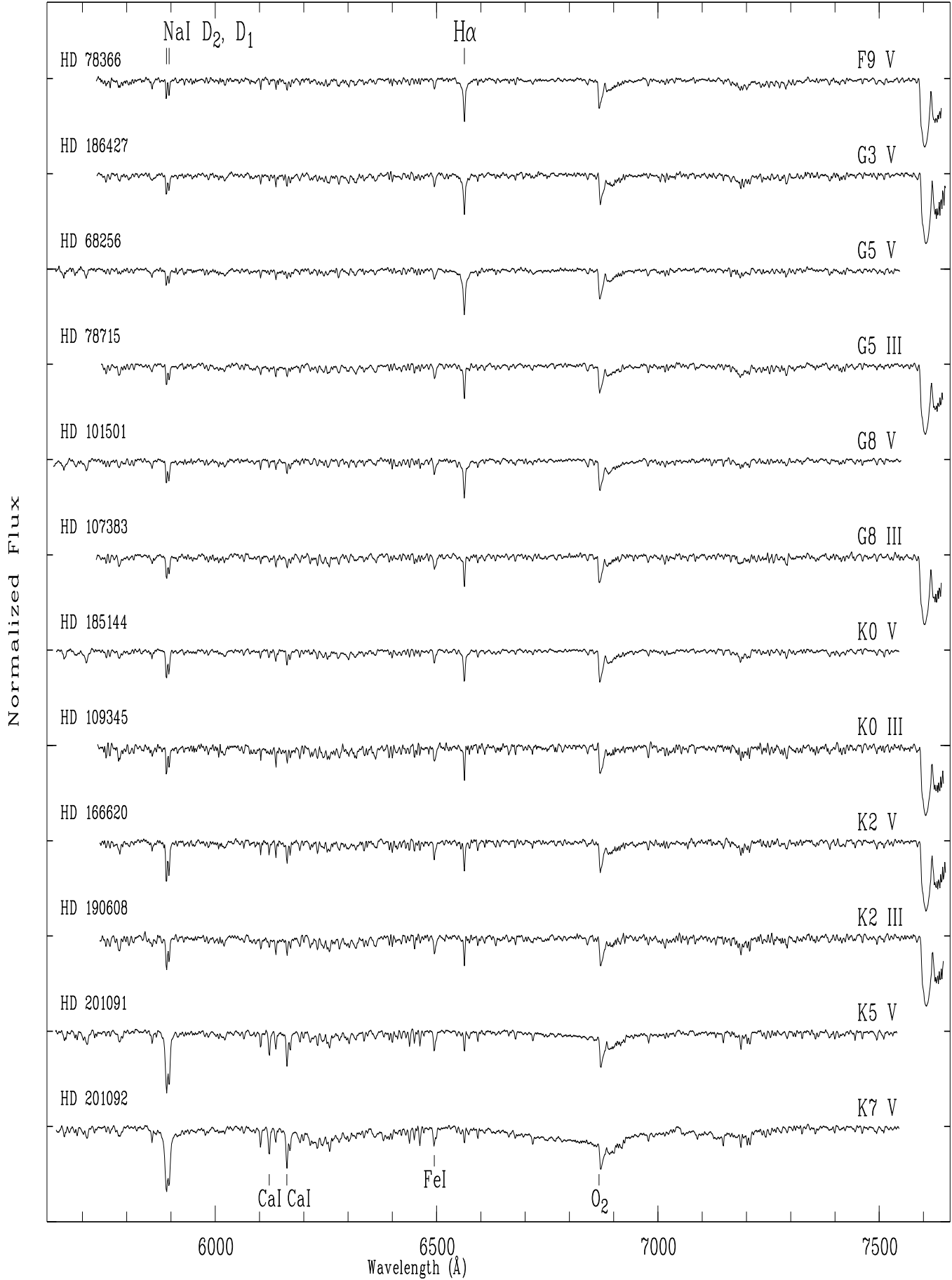,height=37.5cm,width=15.1cm,rheight=25.5cm,rwidth=6cm,angle=270}}
\caption[ ]{Mid-resolution spectra in the
Na~{\sc i} D$_{1}$, D$_{2}$, and H$\alpha$ line region,
in the wavelength range 5650 to 7640 \AA.
\label{fig:mr}}
\end{figure*}

\begin{figure*}
\addtocounter{figure}{-1}
\psscalefirst
\vspace{-0.85cm}
\hspace{-1.6cm}
{\psfig{figure=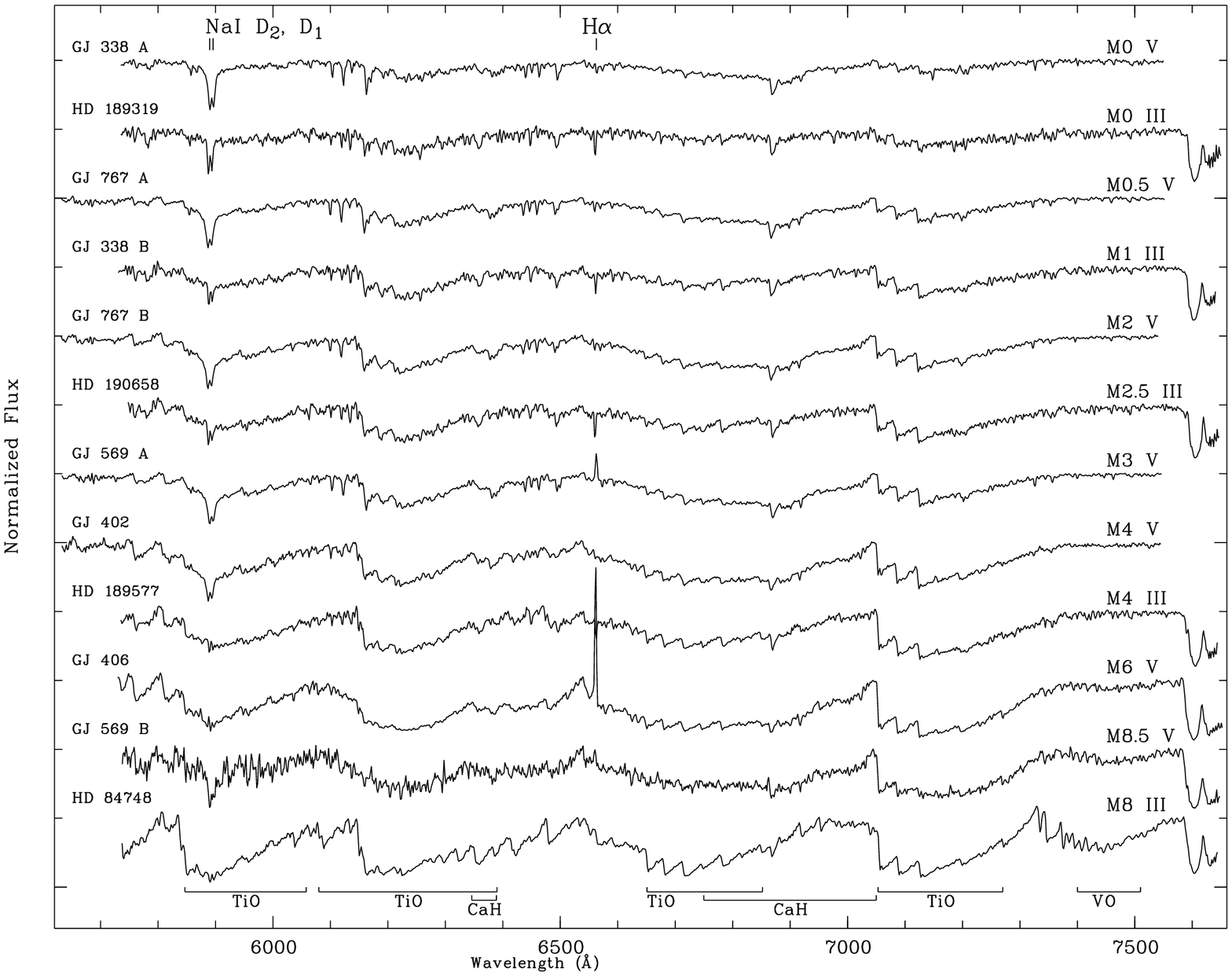,height=37.5cm,width=15.1cm,rheight=25.5cm,rwidth=6cm,angle=270}}
\caption[ ]{continue}
\end{figure*}


\end{document}